\documentclass[aps,pra,groupedaddress,floats,reprint, nofootinbib]{revtex4-2}
\usepackage{graphicx} 
\usepackage[left=20mm, top=10mm, right=20mm, bottom=20mm]{geometry}
\usepackage{physics}
\usepackage{amsmath}
\usepackage{amsfonts}
\usepackage{yfonts}
\usepackage{bm}
\usepackage{xcolor}
\usepackage{hyperref}
\usepackage{subcaption}
\usepackage{ulem}
\usepackage[english]{babel}
\newcommand{\be}{\begin{equation}}
\newcommand{\ee}{\end{equation}}
\newcommand{\bk}{{\bf k}}

\newcommand{\bea}{\begin{eqnarray}}
\newcommand{\eea}{\end{eqnarray}}
\newcommand{\nn}{\nonumber}

\usepackage[sort&compress]{natbib}
\begin{document}
\title{
Generation of vortex electrons by atomic photoionization\\
}

\author{I.\,I.~Pavlov}
\email{ilya.pavlov@metalab.ifmo.ru}
\affiliation{School of Physics and Engineering,
ITMO University, 197101 St. Petersburg, Russia}

\author{A.\,D.~Chaikovskaia}
\email{alisa.katanaeva@metalab.ifmo.ru}
\affiliation{School of Physics and Engineering,
ITMO University, 197101 St. Petersburg, Russia}

\author{D.\,V.~Karlovets}
\email{dmitry.karlovets@metalab.ifmo.ru}
\affiliation{School of Physics and Engineering,
ITMO University, 197101 St. Petersburg, Russia}

\begin{abstract}
We explore the process of orbital angular momentum (OAM) transfer from a twisted light beam to an electron in atomic ionization within the first Born approximation. The characteristics of the ejected electron are studied  regardless of the detection scheme.
We find that the outgoing electron possesses a definite projection of OAM when a single atom is located on the propagation axis of the photon, whereas the size of the electron wave packet is determined solely by the energy of the photon rather than by its transverse coherence length. Shifting the position of the atom yields a finite dispersion of the electron OAM. We also study a more experimentally feasible scenario --- a localized finite-sized atomic target --- and  develop representative approaches to  describing coherent and incoherent regimes of photoionization.
\end{abstract}

\maketitle

\begin{figure}
    {\center
    \includegraphics[width=0.295\linewidth]{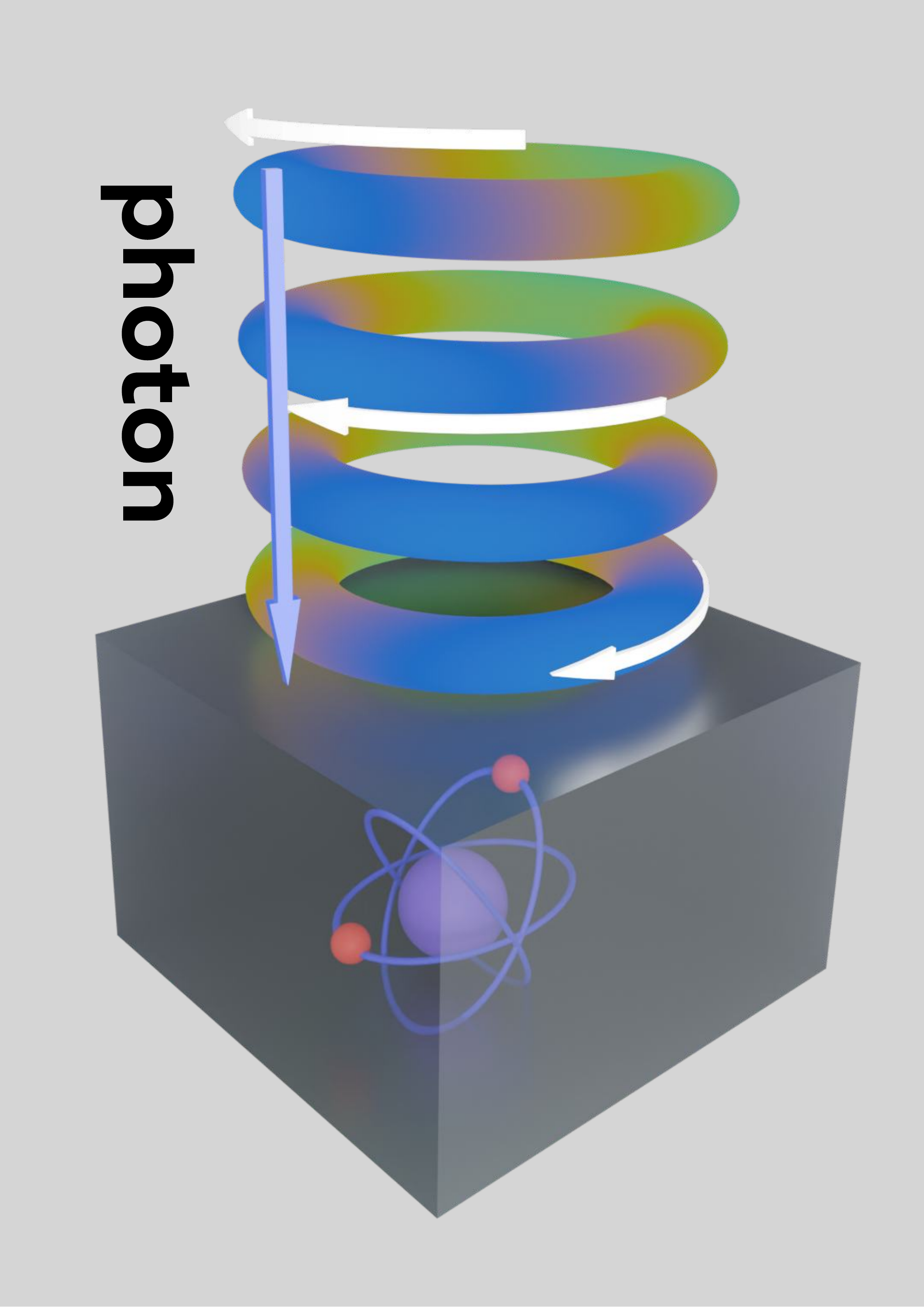}
    \includegraphics[width=0.295\linewidth]{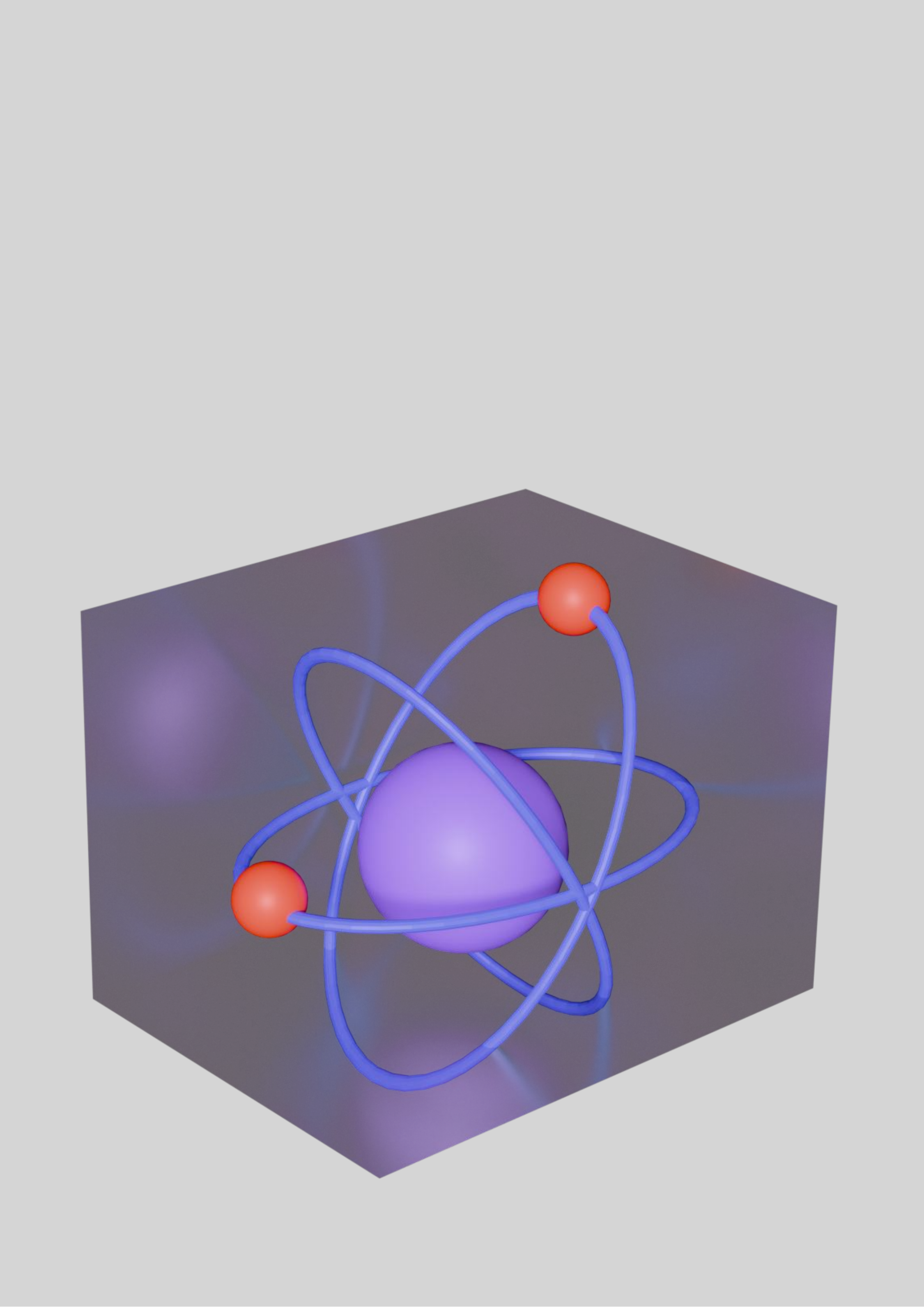}
    \includegraphics[width=0.295\linewidth]{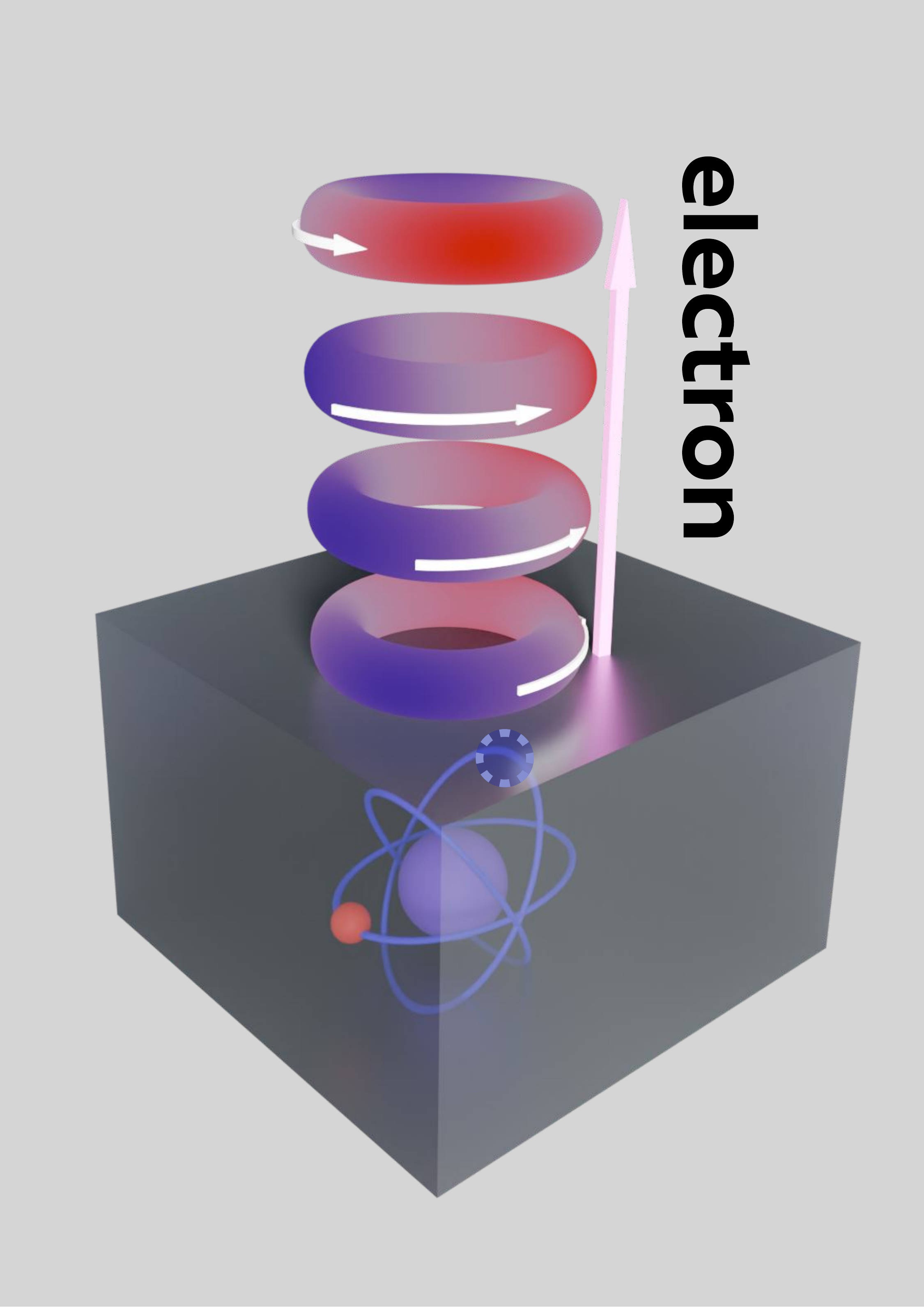}}
    
    \caption{
    Photoionization of atomic target in three stages: {\it left}, twisted light beam descends onto the cathode; {\it central}, the atom(s) in the cathode become excited; {\it right}, a twisted photoelectron is emitted. A cathode can be a single atom or an atomic ensemble.}
    \label{Photoef_pic}
\end{figure}

\textit{Introduction.} 
Atomic photoionization by twisted light, i.e. photons carrying quanta of orbital angular momentum (OAM),  has been previously investigated in Refs.~\cite{matula2013atomic, Surzh_H2+, X_waves_Surzh, surzhykov2016probing, Kiselev_2023, Kiselev_2024, Kosheleva}, with the focus on the process cross section and angular distribution of photoelectrons. However, a fundamental issue {of the angular momentum transfer} remains unaddressed.
In this Letter, we theoretically demonstrate the principal possibility to generate {\it twisted} photoelectrons using the twisted laser beam.
We explore the resulting quantum state of the electron in the photoionization process (see Fig.~\ref{Photoef_pic}), independent of the detection protocol. Particular attention is given to the OAM transfer from the incident photon to the photoelectron. The incident vortex photons are modeled as Bessel beams~\cite{Serbo_UFN} and Laguerre-Gaussian wave packets~\cite{Allen1992}. For the target "cathode", we consider two toy-model options: a single (hydrogen) atom in the $1s$ state and a finite target consisting of symmetrically spread single atoms.

Photoemission of vortex electrons is a possible first step towards generating relativistic electron beams with OAM at particle accelerators. This capability is actively pursued in the joint ITMO-JINR experimental project at the Dzhelepov Laboratory of Nuclear Problems of JINR \cite{RSF_Dubna}.
Relativistic beams of charged particles with OAM offer a unique research tool not only in atomic and molecular physics, diagnostics of nanomaterials, and surface studies --- where orbital momentum provides new sample information --- but also in nuclear physics, spin physics, and hadron physics. Such particles can be used to analyze proton spin, study nuclear forces at low energies, and more \cite{IVANOV2022103987, bliokh2017theory, ivanov2020doing, Karlovets_JHEP, Karlovets_PRD2020}.

Theoretical calculations\footnote{Natural system of units is used throughout the paper with $\hbar = c = 1$ whereas the electron mass and charge are denoted as $m$ and $e < 0$, respectively.} on photoionization commence as a rule with the $S$-matrix element in the first order of perturbation theory \cite{BLP}, {which describes the transition between initial (bound) and final (continuum) electron
states}: 
\begin{equation}
\label{S_fi}
    S_{fi} = 2\pi i \delta(\varepsilon_i+\omega-\varepsilon_f) M_{fi},
\end{equation}
{where $\varepsilon_i$, $\varepsilon_f$ and $\omega$ are the energies of the bound state, continuum state and the photon, respectively.} The transition amplitude $M_{fi}$ can be taken in the non-relativistic form as the energy of the final electron cannot exceed that of the photon, which typically varies from $10$ eV to $100$ eV (optical and ultraviolet range):
\begin{equation}
\label{M_pw_0}
    M_{fi} = -i\frac{e}{m}\int \psi_f^*(\bm{r})\bm{A}(\bm{r}) \cdot \bm{\nabla}\psi_i(\bm{r})\ \dd^3\bm{r}
\end{equation}
with $\psi_i(\bm{r})$ and $\psi_f(\bm{r})$ being the scalar wave functions, and $\bm{A}$ -- the vector potential of a photon.
Employing the first Born approximation is the next obvious step when energies $\omega$ are well above the $1s$ ionization threshold (as is the case with $\varepsilon_i = -1 \operatorname{Ry} = -me^4/2\approx-13.6$ eV for the hydrogen atom). Then, the outgoing electron is described with an asymptotic state with a well-defined momentum, i.e. a plane wave, and such an approach has been successfully applied to describe the ionization by vortex light beams in Refs.~\cite{matula2013atomic, Surzh_H2+, X_waves_Surzh, surzhykov2016probing}. 
A more complex method accounts for the modulation of a free-electron state with the incident laser field by taking the emerging electron wave function in the form of the Volkov-type solution (the so-called Keldysh photoionization theory) \cite{keldysh2024ionization, zheltikov2017keldysh, Faria_2002,  Maxwell}. 

An alternative approach considers the \textit{evolved state} without projecting onto a specific basis \cite{short_EPJC, EPJC}. A somewhat akin method is a nonperturbative approach, in which one numerically solves the exact time-dependent Schrödinger equation, tracking the electron's temporal dynamics without specifying its final state \cite{picon2010_OE, picon2010_NJP, Chen_2023}.

Here, we aim to develop a measurement-strategy independent approach based on robust perturbative calculations to deduce the outgoing photoelectron's quantum state as it evolves from the process on its own. Using the \textit{evolved state} formalism \cite{short_EPJC, EPJC}, we derive the final photoelectron's wave function independently of the post-selection protocol but considering the interaction itself and the initial state. Let $\ket{i}$ be the bound state of the electron; the evolved state $\ket{\text{ev}}$ is linked to it via the $S$-matrix $\hat{S}$. By inserting a complete set of free electron states as plane waves, we obtain the expression
\begin{equation}
\label{ev_0}
    \ket{\text{ev}} = \hat{S}\ket{i} = \int\frac{\dd^3 p_{f}}{(2\pi)^3}\ket{\bm p_{f}} S_{fi}(\bm p_{f}).
\end{equation}
Projection onto a state with a definite momentum $\ket{\bm p}$ yields the final state wave function in the momentum representation:
\bea \nn
    \psi_{\text{ev}}(\bm{p}) \equiv \braket{\bm{p}}{\text{ev}} = \int\frac{\dd^3 p_{f}}{(2\pi)^3}\braket{\bm{p}}{\bm{p}_{f}}S_{fi}(\bm p_{f})\\
    =\int \dd^3 p_{f}\delta(\bm{p}-\bm{p}_{f})S_{fi}(\bm p_{f}) = S_{fi}(\bm{p}).
\eea
{Thus, the wave function of the evolved state in momentum representation simply \textit{coincides with the S-matrix element}. The importance of the S-matrix element \textit{phase}, $\text{arg}\,S_{fi}$, is showcased in this approach, as the processes in particle physics are commonly described with only the absolute value, $|S_{fi}|$.}

Since we describe the electron with the Schrödinger equation and do not take its spin into account, the OAM of the electron coincides with its total angular momentum (TAM). The OAM projection operator in the momentum representation acts as follows
\begin{equation}
    \bra{\bm{p}}\hat{L}_z \ket{\text{ev}} = -i \pdv{\varphi_p}\braket{\bm{p}}{\text{ev}}.
\end{equation}

First, we reproduce the result for the amplitude~\eqref{M_pw_0} in the scenario with the incident photon being a plane wave  with momentum $\bk=(\bm{k}_\perp, k_z)=\omega( \sin{\theta_k}\cos{\varphi_k}, \sin{\theta_k}\sin{\varphi_k},  \cos{\theta_k})$ and helicity  $\Lambda=\pm 1$, it is also convenient to take the Coulomb gauge
\begin{equation}
\label{PWPh}
    A = (0,\bm{A}_{\bm{k}\Lambda}(\bm{r}))
,\ 
    \bm{A}_{\bm{k}\Lambda}(\bm{r}) = \bm{e}_{\bm{k}\Lambda} e^{i\bm{kr}}.
\end{equation}
 From hereon we assume the initial electron to be described by the ground state wave function of the ``hydrogen" atom with charge $Ze$ \cite{BLP}: 
$\psi_i(\bm{r}) = Z^{3/2}e^{-Zr/a}/\sqrt{\pi a^3},$
where $a = 1/(m e^2)$ is the Bohr radius.
The asymptotics of the complete basis of the final electron states can be chosen as plane waves:
$    \psi_f(\bm{r})  = e^{i\bm{p} \bm{r}}, \; \bm{p} =(\bm{p}_\perp, p_z)=p( \sin{\theta_p}\cos{\varphi_p}, \sin{\theta_p}\sin{\varphi_p}, \cos{\theta_p}).$
This constitutes the first Born approximation, in which one neglects the interaction of the final electron with the laser field and the electron–ion attraction after the ionization process. In other words, the electron is assumed to be emitted into the free space. The plane wave amplitude then becomes \cite{matula2013atomic}
\begin{equation}
\label{M_pw}
     M_{fi} = \mathcal{N}p\frac{\bm{n}\bm{e}_{\bm{k}\Lambda}}{\left(\frac{Z^2}{a^2}+q^2\right)^2},
\end{equation}
where $\mathcal{N} = -\frac{8e}{m}\sqrt{\frac{Z^5\pi}{a^5}}$, $\bm{n} = \bm{p}/p$ and $\bm{q} = \bm{p}-\bm{k}$ is the transferred momentum.

\textit{Bessel beam.} 
When the incoming photon is in a {\it twisted state}, a single plane wave is replaced by a superposition of plane waves. Let us consider a Bessel beam and also introduce the possible impact parameter $\bm{b}$:

\begin{align}
\label{Bessel_beam}
    \bm{A}_{\kappa \ell k_z\Lambda}(\bm{r}) &= \int \frac{\dd^2 k_\perp}{(2\pi)^2}a_{\kappa \ell}(\bm{k}_\perp)\bm{A}_{\bm{k}\Lambda}(\bm{r})e^{-i\bm{k}_\perp \bm{b}},\\
    a_{\kappa \ell}(\bm{k}_\perp) &= \sqrt{\frac{2\pi}{\kappa}} (-i)^\ell e^{i\ell \varphi_k}\delta(\abs{\bm{k}_\perp}-\kappa)
\end{align}
with $\kappa$ being the absolute value of the transverse momentum and $\ell=0, \pm 1, \pm 2, \dots$ being the $z$-projection of the TAM. The amplitude, and correspondingly the evolved wave function is obtained through the integration of the plane wave amplitude \eqref{M_pw} with the same weights as in Eq.~\eqref{Bessel_beam}. Then, the amplitude for ionization by the Bessel beam representing the process from Fig.~\ref{Photoef_pic} is

\begin{align} \label{M_tw_0}
    &M_{fi}^{\text{\text{TW}}} =  \int \frac{\dd^2 k_\perp}{(2\pi)^2}a_{\kappa \ell}(\bm{k}_\perp)e^{-i\bm{k}_\perp \bm{b}}M_{fi}\\
\nn    &=\mathcal{N}p\sqrt{\frac{\kappa}{2\pi}} (-i)^\ell  \int \frac{\dd \varphi_k}{2\pi} e^{i\ell \varphi_k - i\kappa b \cos(\varphi_k-\varphi_b)}\frac{\bm{n}\bm{e}_{\bm{k}\Lambda}}{\left(\frac{Z^2}{a^2}+q^2\right)^2}.
\end{align}
Introducing
$\Tilde{\varphi} \equiv  \varphi_k-\varphi_p$ and rewriting $\frac{Z^2}{a^2} + q^2
     = \alpha - \beta \cos{\Tilde{\varphi}}$ in terms of the variables $\alpha \equiv \frac{Z^2}{a^2} + p^2 + k^2 - 2p_z k_z$ and $\beta \equiv 2 p_\perp k_\perp$ 
allows us to put the amplitude as follows (see details in the Supplemental Material~\cite{SUPM}):
\begin{align}
\nn
    &M_{fi}^{\text{TW}} = 
    \mathcal{N}\sqrt{\frac{\kappa}{2\pi}} (-i)^\ell e^{i \ell \varphi_p} \Bigg\{\frac{p_\perp}{\sqrt{2}}\Big[d_{-1\Lambda}^1 I_{\ell+1}(\alpha, \beta, \bm{b}, \varphi_p)\\ 
    &- d_{1\Lambda}^1 I_{\ell-1}(\alpha, \beta, \bm{b}, \varphi_p)\Big] + 
    p_z d_{0\Lambda}^1 I_{\ell}(\alpha, \beta, \bm{b}, \varphi_p)
    \Bigg\}, \label{M_tw}
\end{align}
where $d^J_{MM'}$ are the small Wigner matrices \cite{Varshalovich}, and we define the function
\begin{equation} \label{lIntegral}
     I_{\ell}(\alpha, \beta, \bm{b}, \varphi_p) \equiv \int \frac{\dd \Tilde{\varphi}}{2\pi} e^{i\ell \tilde{\varphi} - i\kappa b \cos (\Tilde{\varphi}+\varphi_p-\varphi_b)}\frac{1}{(\alpha - \beta \cos{\Tilde{\varphi}})^2}.
\end{equation}

In fact, we can make an immediate conclusion about the evolved state of the photoelectron when the ionizing beam falls onto the target atom with a vanishing impact parameter. If $\bm{b}=0$, $I_{\ell}(\alpha, \beta, 0)$ does not depend on $\varphi_p$. Thus, 
\begin{equation}
    -i \pdv{\varphi_p} M_{fi}^{\text{TW}} = \ell  M_{fi}^{\text{TW}},
\end{equation} leading to  $-i \pdv{\varphi_p} S_{fi}^{\text{TW}} = \ell  S_{fi}^{\text{TW}}$
and, consequently,
\begin{equation}
   \hat{L}_z \ket{\text{ev}} = \ell\ket{\text{ev}}.
\end{equation}
This equation illustrates that the evolved state represents a vortex electron with the OAM $z$-projection equal to $\ell$. {A similar result can be expected for the ionization of some excited state of the hydrogen-like atom with a well-defined OAM projection $\ell_e$ onto the $z$ axis, as the conservation of the OAM projection is a consequence of the system being axially symmetric. In this case the electron OAM would become $\ell+\ell_e$. }
Moreover, for $\bm{b}=0$, the amplitude can be evaluated analytically as \cite{Karlovets2017, serbo2015} 
\begin{equation}
    I_{\ell}(\alpha, \beta, 0) = -\pdv{\alpha} \left\{\left(\frac{\beta}{\alpha + \sqrt{\alpha^2-\beta^2}}\right)^{|\ell|}\frac{1}{\sqrt{\alpha^2-\beta^2}}\right\}. 
\end{equation}

 For a non-vanishing impact parameter, we make an estimation supposing that $\kappa b$ is a reasonable small parameter in our problem. For realistic wavelengths of the laser field $\lambda \sim 10^2-10^3 $ nm and the Bessel beam opening angle $\theta_k \sim 1^\circ$ it means that we consider values $b \ll 1-10$ $\mu$m, which are relevant to the beam axis positioning accuracy in experiments with trapped atoms \cite{PenningTrap2016, schmiegelow2016transfer, PhysRevLett.129.263603}. 
Expansion of the impact parameter - dependent exponent in \eqref{lIntegral} up to the linear term:
\bea \nn
    e^{- i\kappa b \cos (\Tilde{\varphi}+\varphi_p-\varphi_b)} \approx & \ 1 - \frac{i \kappa b}{2}e^{-i \varphi_b} e^{i(\Tilde{\varphi}+\varphi_p)}\\
    &- \frac{i \kappa b}{2}e^{i \varphi_b} e^{-i(\Tilde{\varphi}+\varphi_p)}
\eea
is equivalent to making the following replacement in Eq.~\eqref{M_tw}:
\bea \nn 
     I_{\ell}(\alpha, \beta, \bm{b}) \rightarrow & I_{n}(\alpha, \beta, 0) - \frac{i \kappa b}{2}e^{-i \varphi_b} e^{i\varphi_p} I_{\ell+1}(\alpha, \beta, 0) \\
    & - 
      \frac{i \kappa b}{2}e^{i \varphi_b} e^{-i\varphi_p} I_{\ell-1}(\alpha, \beta, 0). \label{I_series}
\eea
We see that in this case the evolved state is a superposition of the twisted states with OAM values $l$ and $l\pm 1$ rather than an eigenstate of $\hat{L}_z$.
Analogously, the expansion of the exponent up to the arbitrary $n$-th term gives rise to the new terms in the superposition with OAM $l\pm n$. The coefficients at these terms are suppressed by the factor of $(\kappa b)^n$. The photoelectron not representing a single twisted state is unsurprising, as the displacement of the target atom from the wavefront center breaks the system's cylindrical symmetry.
\begin{figure*}[!htb]
        \centering
	\begin{subfigure}{0.32\linewidth}
        \center
		\includegraphics[width=\linewidth]{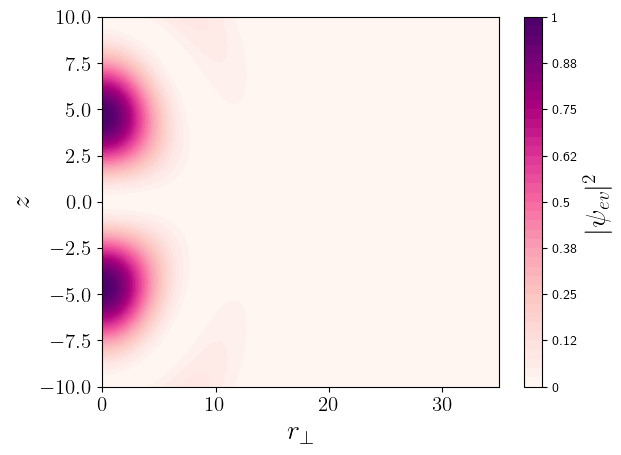}\\(a)
	\end{subfigure}
    \begin{subfigure}{0.32\linewidth}
        \center
		\includegraphics[width=\linewidth]{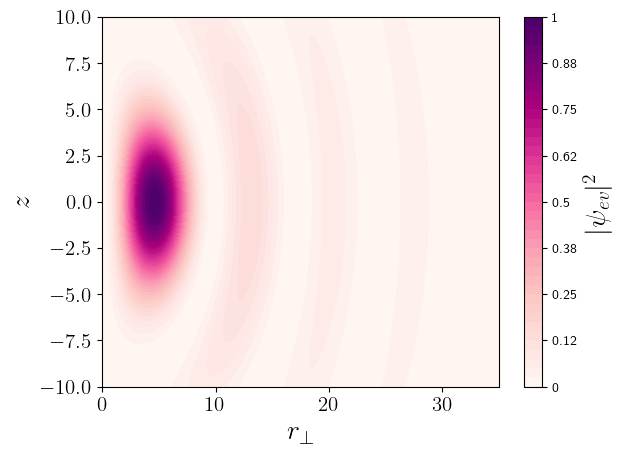}\\(b)
	\end{subfigure}
    \begin{subfigure}{0.32\linewidth}
        \center
		\includegraphics[width=\linewidth]{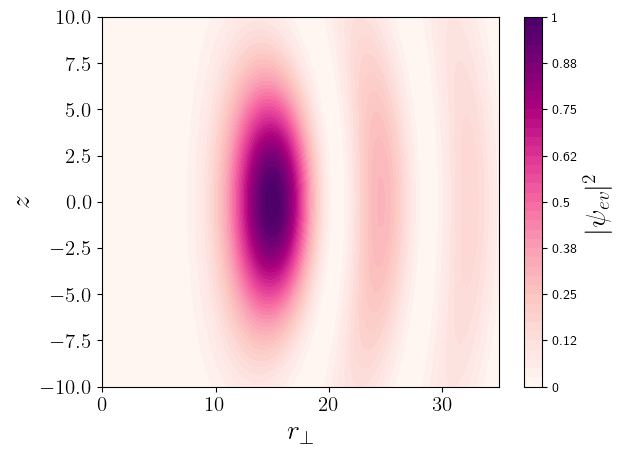}\\(c)
	\end{subfigure}
 
	\caption{Normalized probability density $|\psi_{\text{ev}}(r_\perp, z, \varphi=0)|^2$ of the evolved state. Due to the azimuthal symmetry of the problem there is no dependence on $\varphi$. The incident photon represents Bessel beam with $\omega = 1.2|\varepsilon_i|$, {$\Lambda$ = 1,} opening angle $\theta_k = \arctan(\kappa / k_z) = 1^\circ$. Length is measured in Bohr radii $a$. (a) $\ell = 0$; (b) $\ell = 1$; (c) $\ell = 5$. For larger energies of the photon the plots are qualitatively the same with the exception that the probability density is more localized in all directions.}
	\label{Prob_dens_Bess}
\end{figure*}

 

\textit{Laguerre-Gaussian beam.}
Let us now consider a more realistic model of a twisted photon beam, represented by a Laguerre-Gaussian (LG) wave packet with \textit{OAM} projection $\ell$ and helicity $\Lambda$. Analogously to Eq.~\eqref{M_tw_0}, the photoionization amplitude with an incident LG beam is given by
\begin{align}
\label{M_LG}
    M_{fi}^{LG} =& \displaystyle \int \frac{\dd^2 k_\perp}{(2\pi)^2}U(\bm{k}_\perp)e^{i\Lambda \varphi_k-i\bm{k}_\perp \bm{b}}M_{fi} \\
    =&p\mathcal{N} \int \frac{\dd^2 k_\perp}{(2\pi)^2}U(\bm{k}_\perp)e^{i\Lambda \varphi_k-i\bm{k}_\perp \bm{b}} \frac{\bm{n}\bm{e}_{\bm{k}\Lambda} }{\left(\frac{Z^2}{a^2}+q^2\right)^2}, \nn
\end{align}
with the function 
\begin{align}\nn
     U(\bm{k}_\perp) = &(-1)^p  2^{-l/2} i^{-\ell}  \pi w_0^{\ell + 1} k_\perp^\ell \exp\left[-\frac{k_\perp^2w_0^2}{4}\right]\\ &\times L_p^\ell\left(\frac{k_\perp^2 w_0^2}{2}\right)\exp\left[i\ell \varphi_k\right].
\end{align}
{Here, $L_p^\ell$ refers to the associated Laguerre polynomial and $w_0$ to the  beam waist.} The details of the derivation are presented in the Supplemental Material~\cite{SUPM} (see also references \cite{andrews2012angular, Peshkov2017} therein).

Evaluating the transition amplitude \eqref{M_LG} is more cumbersome than for the Bessel beam, as the integral over $k_\perp$ appears in the expression and $k_z$ becomes a function of $k_\perp$. However, the result remains valid for a vanishing impact parameter, showing that the electron's evolved state is twisted:
\begin{equation}
    -i \pdv{\varphi_p} M_{fi}^{LG} = (\ell +\Lambda) M_{fi}^{LG}
\end{equation}
Previous results on the OAM superposition in the photoelectron state for a non-vanishing impact parameters also remain valid for the LG beam.


\textit{Transverse coherence length of the photoelectron.} With the outgoing electron's wave function in momentum representation derived, we now evaluate its Fourier transform to investigate the probability density in coordinate space. First, we rewrite the delta function in \eqref{S_fi} as
\begin{equation}
    \delta\left(\varepsilon_i + \omega - \frac{p^2}{2m}\right) =  \frac{m}{p_\perp}\delta\left(p_\perp - \sqrt{2m(\varepsilon_i + \omega) - p_z^2}\right)
\end{equation}
Then for the Bessel photon and zero impact parameter, the Fourier integral becomes
\begin{align}
\nn
    \psi_{\text{ev}}(\bm r) = 
     e^{i\ell\varphi} \mathcal{N}m \sqrt{\frac{\kappa}{(2\pi)^5}} \int_{-\sqrt{2m(\varepsilon_i+\omega)}}^{\sqrt{2m(\varepsilon_i+\omega)}} J_\ell(p_\perp r_\perp)\\ \nn
     \times \Bigg[\frac{p_\perp}{\sqrt{2}}\big[d_{-1\Lambda}^1 I_{\ell+1}(\alpha, \beta, 0) - d_{1\Lambda}^1 I_{\ell-1}(\alpha, \beta, 0)\big]\\ \label{psi(r)}
     + p_z d_{0\Lambda}^1 I_{\ell}(\alpha, \beta, 0)
    \Bigg] e^{ip_z z}dp_z,
\end{align}
where $p_\perp$, $\alpha$ and $\beta$ are the functions of the integration variable $p_z$. {Here $J_\ell$ is the Bessel function and $r_\perp$ is the radial distance from the symmetry axis.} The remaining integral on $p_z$ in Eq.~\eqref{psi(r)} can be computed numerically. For the incident LG photon the Fourier transform is done analogously with an additional numerical integration over $k_\perp$. 

Fig.~\ref{Prob_dens_Bess} shows the probability density $|\psi_{\text{ev}}(\bm r)|^2$ of the photoelectron for a Bessel beam with different values of OAM projection. Since we focus on the spatial distribution, the absolute value of the probability density is normalized such that its maximum value equals $1$. At zero impact parameter, the probability density is azimuthally symmetric, so we present its dependence only on the radial coordinate $r_\perp$ and $z$.
The photoelectron wave packet is localized around the atom's position in both transverse and longitudinal directions. For a Bessel beam with zero OAM, the maximum of the probability density lies on the $z$ axis. For non-zero values of $\ell$, the wave packet exhibits a characteristic "doughnut shape" typical of vortex states. Additionally, the radius of the first ring increases with higher values of $\ell$.

In the Supplemental Material~\cite{SUPM} we also demonstrate the probability density of the photoelectron for an incident LG beam with different values of the radial index $p$ and OAM $\ell$. The distributions for both types of beams with the same value of OAM are nearly identical despite the fundamental difference between Bessel and LG photons: the latter is localized in the transverse plane while the former is not. This similarity illustrates that the width of the photoelectron wave packet is independent of the transverse coherence of the incident light. This feature can be explained by Eq.~\eqref{psi(r)}, where the transverse coordinate $r_\perp$ appears only in the argument of the Bessel function $J_\ell(p_\perp r_\perp)$, alongside the transverse momentum $p_\perp = \sqrt{2m(\varepsilon_i+\omega) - p_z^2}$, which does not explicitly depend on $k_\perp$. Therefore, we conclude that the width of the photoelectron wave packet is determined by the photon energy $\omega$ and not its transverse momentum $\kappa$.

\begin{figure}[!htb]
\centering
	\includegraphics[width=\linewidth]{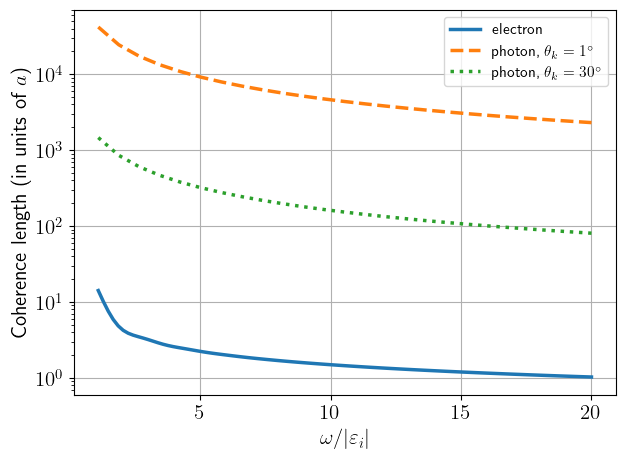}
	\caption{Comparison of transverse coherence length of the photoelectron (solid blue line) and characteristic size of the Bessel beam $\ell / \kappa$ for $\ell=3$ depending on the photon energy $\omega$ for the Bessel beam opening angles $\theta_k = 1^\circ$ (dashed orange line) and $\theta_k = 30^\circ$ (dotted green line). The width of the electron wave packet does not depend on the photon opening angle. }
 \label{Coh_length}
\end{figure}

Let us now explicitly define the width (\textit{the transverse coherence length}) of the electron packet as the distance from the $z$ axis to the point where the probability density is maximal. Further, we demonstrate the dependence of the electron transverse coherence length on the energy of the incident photon and compare it with the characteristic radius of the Bessel beam $\ell/\kappa$.
As seen from Fig. \ref{Coh_length}, the width of the electron packet, which is many orders of magnitude smaller than the Bessel beam radius, decreases with the increase of the photon energy. For $\omega \sim 20 |\varepsilon_i|$ the localization becomes comparable to the Bohr radius $a$, which is the initial localization distance in the atom. The unbounded growth of the width as $\omega \to |\varepsilon_i|$ has little physical sense since the Born approximation that we use becomes invalid in case of extremely small energies of the outgoing electron.

\textit{Mesoscopic target.} Up to this point, we have discussed the photoionization of a single atom positioned at a specific impact parameter relative to the propagation axis of the vortex beam. While recent advancements have made it possible to localize the target in this way \cite{schmiegelow2016transfer,PhysRevLett.129.263603}, the majority of photoionization experiments involve extended (mesoscopic) targets.

It is generally acknowledged in related literature that there is no coherence between photoelectrons emitted from different atoms \cite{chambers1992elastic}. Thus, describing photoemission from localized targets involves averaging the cross section over a target distribution function \cite{X_waves_Surzh, surzhykov2016probing, Surzh_H2+}. However, coherent effects in photoionization are also recognized. Numerous theoretical and experimental works examine photoemission from diatomic molecules, akin to Young’s double-slit experiment. Here, the coherent emission of electrons from spatially separated sources leads to interference patterns in the differential cross-section \cite{Fano1966, baltenkov2012interference, kunitski2019double, akoury2007simplest, Surzh_H2+}.

A peculiar feature of the evolved state arises in the hypothetical scenario when the photoionization occurs coherently from all atoms in the target which have the same phase of the wave function. It means that it is the \textit{amplitude} of the process that should be averaged over the target rather than its squared absolute value (cross-section):
\begin{equation}
    \widetilde{M}_{fi}^{\text{TW}} = \int \dd^2 b \:f(\bm b) M_{fi}^{\text{TW}}   = \int b\:\dd b \:\dd \varphi_b  \:f(\bm{b}) M_{fi}^{\text{TW}}.
\end{equation}
{Here $f(\bm b)$ is the two-dimensional distribution function of the target.}

Let us, in addition, assume the mesoscopic target to be \textit{axially symmetric}.
One can notice that in this case the integration over $\varphi_b$ eliminates any term containing the factor $e^{i n \varphi_b}$ in the amplitude \eqref{M_tw} {(The details are presented in the Supplemental Material~\cite{SUPM})}. Thus, after the averaging over such a target the evolved state becomes a single-mode twisted state, just as in the scenario with $\bm{b}=0$. {From an experimental viewpoint such a scenario could corresponds, for example, to the target represented by a cold atomic gas, where the coherence effects can play an important role \cite{UES_proposal, ColdGas}.}

\textit{Incoherent photoionization: density matrix approach.} Let us consider the density matrix of the (pure) evolved state
\begin{equation}
    \hat{\rho}_{\text{ev}} = \ket{\text{ev}}\bra{\text{ev}}.
\end{equation}
Unlike the regular density matrices, $
    \Tr (\hat \rho) \neq 1,$
as
$    \Tr (\hat \rho_{\text{ev}}) = \bra{\text{ev}}\ket{\text{ev}} = \int \frac{\dd^3 p}{(2\pi)^3} |\psi_{\text{ev}}(\bm p)|^2$, which diverges. 
Thus, if we would like to evaluate the mean values of observables using this density matrix, we should take the normalized definition: $    \expval{\hat A} = \frac{\Tr\left(\hat{A} \hat \rho_{\text{ev}}\right)}{\Tr \left(\hat \rho_{\text{ev}} \right)}$.
In the Supplemental Material we check that thus defined average value of the OAM projection upon the $z$ axis calculated with this density matrix equals $\ell$, $\expval{\hat L_z}=\ell$, when the impact parameter equals zero.

Analogously, for a non-vanishing impact parameter we have
\begin{widetext}
\begin{align}
\nn
    \expval{\hat L_z} = &\frac{1}{\Tr(\hat \rho_{\text{ev}})} \sum_{\tilde \ell}\tilde \ell \int \frac{q_\perp \dd  q_\perp \dd  q_z \dd \varphi_q \dd \varphi_q'}{(2\pi)^6}  (2\pi)^2 e^{i\tilde \ell (\varphi_q - \varphi_q')} \psi_{\text{ev}}^*(q_\perp, q_z, \varphi_q, \bm b) \psi_{\text{ev}}(q_\perp, q_z,\varphi_q', \bm b) \\ \nn
    &= 2\pi \mathcal{N}^2 p^2 \kappa \sum_{\tilde \ell}\tilde \ell \int \frac{q_\perp \dd  q_\perp \dd  q_z \dd \varphi_q \dd \varphi_q'}{(2\pi)^6}  (2\pi)^2 e^{i(\tilde \ell - \ell)\varphi_q} e^{-i(\tilde \ell - \ell)\varphi_q'} \delta\left(\varepsilon_i + \omega - \frac{q^2}{2m}\right) \\ \nn
    &\times \delta\left(\varepsilon_i + \omega - \frac{q'^2}{2m}\right) \Bigg[\frac{q_\perp}{\sqrt{2}}\left[d_{-1\Lambda}^1 I_{\ell+1}(\alpha, \beta, \bm{b}, \varphi_q) - d_{1\Lambda}^1 I_{\ell-1}(\alpha, \beta, \bm{b}, \varphi_q)\right] + 
    q_z d_{0\Lambda}^1 I_{\ell}(\alpha, \beta, \bm{b}, \varphi_q)\Bigg]^* \\ \label{expv L_z via rho}
    &\times \Bigg[\frac{q_\perp'}{\sqrt{2}}\left[d_{-1\Lambda}^1 I_{\ell+1}(\alpha', \beta', \bm{b}, \varphi_q') - d_{1\Lambda}^1 I_{\ell-1}(\alpha', \beta', \bm{b}, \varphi_q')\right] + 
    q_z' d_{0\Lambda}^1 I_{\ell}(\alpha', \beta', \bm{b}, \varphi_q') \Bigg]\bigg/{\Tr(\hat \rho_{\text{ev}})}.
\end{align}

\end{widetext}

Let us again suppose that the target is small enough and expand expression~\eqref{expv L_z via rho} in powers of $\kappa b$ up to the linear terms. Recalling expression~\eqref{I_series},
we observe that any term in Eq.~\eqref{expv L_z via rho} proportional to $\kappa b$ has the form of
$\sim I_n(\alpha, \beta, 0) I_m(\alpha', \beta', 0) e^{\pm i (\varphi_b - \varphi_q)}$
or $\sim I_n(\alpha, \beta, 0) I_m(\alpha', \beta', 0) e^{\pm i (\varphi_b - \varphi_q')}$.
After the integration over $\varphi_q$  (or $\varphi_q'$) in Eq.~\eqref{expv L_z via rho} the Kronecker delta $\delta_{\tilde \ell, \ell\pm 1}$ appears. Then this term is eliminated by the second azimuthal integral, as

\begin{equation}
    \int_0^{2\pi} e^{i(\tilde \ell - \ell)\varphi_q} = 0
\end{equation}
when $\tilde \ell = \ell \pm 1$.
Thus, we come to the conclusion that in the first order expansion in the parameter $\kappa b$ 
\begin{equation}
    \expval{\hat L_z} = \ell
\end{equation}
despite the small, but non-vanishing impact parameter.



Importantly, the \textit{incoherent} averaging over the target, which is generally understood as the averaging of the differential cross-section, can be interpreted in terms of the density matrix. In this case the outgoing electron simply represents a \textit{mixed state} of electron states scattered from different atoms  rather than a coherent superposition. We now introduce the averaged density matrix
\begin{equation}
   \hat{\rho}_{\text{ev}}' = \int \dd^2 b \:f(\bm b) \hat{\rho}_{\text{ev}}(\bm b).
\end{equation}
If the target is small enough in the sense that $\kappa b \ll 1$, the mean value of the OAM projection of each state in the mixture approximately equals $\ell$, and thus for the whole state we find that
\begin{equation}
    \expval{\hat L_z}' = \frac{1}{\Tr(\hat \rho_{\text{ev}}')}\Tr\left(\hat L_z \hat \rho_{\text{ev}}' \right) =\ell.
\end{equation}
{Although the described photoelectron state is not pure, it may be deemed 
twisted in the sense of having a definite expectation value of the OAM projection.}

\textit{Conclusion.}
In summary, we have theoretically studied the final evolved state of an electron emitted in the ionization of a hydrogen-like atom by a vortex light beam. Non-relativistic first-order perturbation theory was used to derive the photoelectron's state as it evolves from the process itself, independent of any detection scheme. We have demonstrated that the emitted photoelectron generally appears as a localized wave packet with a well-defined projection of orbital angular momentum (OAM) onto the quantization axis of the photon total angular momentum (TAM).

The transverse size (coherence length) of the electron packet is solely governed by the energies of the initial bound electron and the incident photon. For ionization of an extended atomic target, as long as the target is axially symmetric and its size does not exceed the transverse size of the light beam, the OAM of the resulting electron is not significantly violated.

While we have focused on non-relativistic electrons and neglected spin effects, it is expected that in a rigorous relativistic framework, angular momentum is fully transferred from the photon to the electron. However, in this case, the outgoing particle has a specific value of TAM projection rather than OAM, and the separation of TAM into spin and orbital components is ambiguous for both particles \cite{Bauke}.

Therefore, we propose that ionization by vortex light beams can be used to generate vortex electrons in experimental setups where the usage of other generation techniques (diffraction grating, phase plate, magnetic monopole, etc. \cite{bliokh2017theory}) is hampered for some reason. For instance, such an approach could be used to produce high-energy twisted electrons via irradiating photocathodes 
at various accelerator facilities, such as LINAC-200 at JINR, Dubna. This could open up opportunities for the experimental study of vortex state collisions and various radiation processes \cite{ivanov2020doing, short_EPJC, EPJC, IVANOV2022103987, Ivanov_Ultarel, Chaikovskaya,  Sheremet_eLG, Karlovets_JHEP, Karlovets_PRD2020}. 

{More results related to the ITMO-JINR project on relativistic vortex electron generation could be expected in near future. In particular, we plan to take into account the possibility of OAM redistribution between the photoelectron and the atomic center of mass (similarly to \cite{PhysRevLett.129.263603, Surzhykov_interplay_CM, Baturin2024_atom}), and the processes of relaxation due to interaction with the adjacent atoms.}

\textit{Acknowledgments.} 
We thank V. Ivanov for his initial contribution and G. Sizykh and D. Grosman for valuable discussions and advice. The  studies with the Bessel beam are supported by the Ministry of Science and Higher Education of the Russian Federation (agreement No. 075-15-2021-1349). Research on photoionization with LG beams is supported by the Russian Science Foundation (Project No. 23-62-10026; \cite{RSF_Dubna}). A.C. and D.K.'s work on incoherent photoionization is supported by the ITMO Fellowship and Professorship Program. Analysis of the transverse coherence length is also supported by the Foundation for the Advancement of Theoretical Physics and Mathematics “BASIS”.

\bibliography{ref}

\end{document}


\title{Supplemental Material to "Generation of vortex electrons by atomic photoionization"}

\author{I.\,I.~Pavlov}

\author{A.\,D.~Chaikovskaia}

\author{D.\,V.~Karlovets}

 \maketitle

\section{Polarization vector decomposition}
\label{app_polar_def}
Let us take a plane wave photon 
 with momentum $\bk=(\bm{k}_\perp, k_z)=\omega( \sin{\theta_k}\cos{\varphi_k}, \sin{\theta_k}\sin{\varphi_k},  \cos{\theta_k})$ and helicity  $\Lambda=\pm 1$ in representation
\begin{equation}
\label{PWPh}
    A = (0,\bm{A}_{\bm{k}\Lambda}(\bm{r}))
,\ 
    \bm{A}_{\bm{k}\Lambda}(\bm{r}) = \bm{e}_{\bm{k}\Lambda} e^{i\bm{kr}}.
\end{equation}
The polarization vector $\bm{e}_{\bm{k}\Lambda}$ can be written
in terms of spherical angles $\theta_k$ and $\varphi_k$ \cite{Serbo_UFN}
\begin{equation}
\label{e_klambda}
    \bm{e}_{\bm{k}\Lambda} = \sum\limits_{\sigma=0,\pm 1}e^{-i\sigma\varphi_k}d_{\sigma\Lambda}^1 (\theta_k)\bm{\chi}_\sigma,
\end{equation}
where $d^J_{MM'}(\theta)$  are the small Wigner functions \cite{Varshalovich} and the basis vectors are defined as follows:
\begin{align}  
&    \bm{\chi}_0 = \begin{pmatrix}
        0 \\
        0 \\
        1
    \end{pmatrix}
    ,\, 
    \bm{\chi}_{\pm 1} = \mp \frac{1}{\sqrt{2}}\begin{pmatrix}
        1 \\
        \pm i \\
        0
    \end{pmatrix};\\
& \bm{\chi}_{1}\bm{\chi}_{-1}=\bm{\chi}_{-1}\bm{\chi}_{1} = -1
,\ 
    \bm{\chi}_{\pm1}\bm{\chi}_{\pm1} = 0.
\end{align}
Further, we can express the vector $\bm{n}=\bm{p}/p$ in the same basis as polarization vector $\bm{e}_{\bm{k}\Lambda}$:

\bea
    &\bm{n} = \bm{\chi}_0 \cos{\theta_p} -\frac{1}{\sqrt{2}}(\bm{\chi}_1 e^{-i\varphi_p} -\bm{\chi}_{-1}e^{i\varphi_p}) \sin{\theta_p}.
\eea
Then, the scalar product in the amplitude for photoionization given be Eq.~$(10)$ can be decomposed as follows:
\be
    \bm{e}_{\bm{k}\Lambda}\bm{n} = \sum\limits_{\sigma=0,\pm 1}e^{-i\sigma\varphi_k}d_{\sigma\Lambda}^1(\theta_k)\bm{\chi}_\sigma \bm{n} = 
    \frac{\sin \theta_p}{\sqrt{2}}e^{i(\varphi_k-\varphi_p)}d_{-1\Lambda}^1(\theta_k)+ d_{0\Lambda}^1(\theta_k)\cos \theta_p -\frac{\sin \theta_p}{\sqrt{2}}e^{-i(\varphi_k-\varphi_p)}d_{1\Lambda}^1(\theta_k).
\ee
From this we can derive Eq.~$(11)$ and proceed with analysis of the evolved state  orbital angular momentum (OAM).

\section{Laguerre-Gaussian beam}
Let us consider a Laguerre-Gaussian (LG) wave packet. The vector potential is taken in the form
\begin{equation}
\label{A_LG}
    \bm{A}(\bm{r}) = \bm{e}_{\Lambda} u(\bm{r}) e^{ikz},
\end{equation}
where the polarization vector $\bm{e}_{\Lambda}$ is supposed to be orthogonal to the $z$ axis and the scalar amplitude $u(\bm{r})$ satisfies the \textit{paraxial} Helmholtz equation \cite{andrews2012angular}. For the LG beam it is chosen as \cite{Allen1992}
\begin{align}
&u\left(r_{\perp}, \phi, z\right)=  \frac{1}{w(z)}\left(\frac{\sqrt{2} r_{\perp}}{w(z)}\right)^\ell  \exp \left[-\frac{r_{\perp}^2}{w^2(z)}\right] L_p^\ell\left(\frac{2 r_{\perp}^2}{w^2(z)}\right) \\ \nn
& \times \exp \left[i \ell \varphi+\frac{i k r_{\perp}^2 z}{2\left(z^2+z_R^2\right)} -i(2 p+\ell + 1) \arctan \left(\frac{z}{z_R}\right)\right].
\end{align}
The beam is \textit{monochromatic} and its TAM projection is $l+\Lambda$. {Here $L_p^\ell$ is the associated Laguerre polynomial and $w_0$ is the so-called beam waist, which determines the Rayleigh range $z_R = kw_0^2/2$ and the beam width $w(z) = w_0\sqrt{1 + z^2/z^2_R}$.}
The atom is supposed to lie at the beam focus ($z=0$), so we can focus on the modifications in the transverse plane. There, the LG profile has a characteristic ringlike pattern  with a finite number of rings determined by the radial index $p$ in contrast to Bessel beams with an infinite number of rings.

It is convenient to also represent the LG beam as a superposition of plane waves:
 \begin{equation}
 \label{A_LG_fourier}
     \bm{A}(\bm{r}) =\int \frac{\dd^2 k_\perp}{(2\pi)^2} U(\bm{k}_\perp) \bm{e}_{\Lambda} e^{i\bm{k}\bm{r}}
 \end{equation}
The modulus of the wave vector is kept fixed: $k = \sqrt{k_z^2 + k_\perp^2} = \operatorname{const}$. The function $U(\bm{k}_\perp)$ is connected to the amplitude $u(\bm{r})$ in the beam focus by the two-dimensional Fourier transform:
\begin{align}
    U(\bm{k}_\perp) = & \displaystyle\int u(r_\perp, \varphi, z=0) e^{-i \bm{k}_\perp \bm{r}_\perp} \dd^2 r_\perp,\\ \nn
    U(\bm{k}_\perp) = &(-1)^p 2^{-l/2} i^{-\ell}  \pi w_0^{\ell + 1} k_\perp^\ell \exp\left[-\frac{k_\perp^2w_0^2}{4}\right] L_p^\ell\left(\frac{k_\perp^2 w_0^2}{2}\right)\exp\left[i\ell \varphi_k\right],
    \label{U_LG}
\end{align}
where the beam waist $w_0 \equiv w(z=0)$.
However, such a vector potential does not satisfy Coulomb gauge, because the superposition contains nontransverse plane waves. To fix this, we make the following modification of the vector potential (see details in \cite{Peshkov2017}):
\begin{equation}
 \label{A_LG_C}
     \bm{A}_{\text{trans}}(\bm{r}) = \int \frac{\dd^2 k_\perp}{(2\pi)^2} U(\bm{k}_\perp) e^{i\Lambda \varphi_k} \bm{e}_{\bm{k} \Lambda} e^{i\bm{k}\bm{r}}
 \end{equation}
with the polarization vector defined in Eq. \eqref{e_klambda}.

Analogously to the case of Bessel beam, the amplitude of the photoionization process with an incident LG beam is given by
\begin{align}
\label{M_LG}
    M_{fi}^{LG} =& \displaystyle \int \frac{\dd^2 k_\perp}{(2\pi)^2}U(\bm{k}_\perp)e^{i\Lambda \varphi_k-i\bm{k}_\perp \bm{b}}M_{fi} \\
    =&p\mathcal{N} \int \frac{\dd^2 k_\perp}{(2\pi)^2}U(\bm{k}_\perp)e^{i\Lambda \varphi_k-i\bm{k}_\perp \bm{b}} \frac{\bm{n}\bm{e}_{\bm{k}\Lambda} }{\left(\frac{Z^2}{a^2}+q^2\right)^2}. 
\end{align}

If the atom is placed not in the focal plane ($z\neq 0$) of the LG beam, every plane wave in the superposition \eqref{A_LG_C} should be multiplied by the corresponding phase factor:

\be
    M_{fi}^{LG} = 
    p\mathcal{N} \int \frac{\dd^2 k_\perp}{(2\pi)^2}U(\bm{k}_\perp)e^{-i\bm{k}_\perp \bm{b} - ib_z\sqrt{k^2-k_\perp^2}} \frac{\bm{n}\bm{e}_{\bm{k}\Lambda} }{\left(\frac{Z^2}{a^2}+q^2\right)^2} 
    \approx 
    p\mathcal{N} \int \frac{\dd^2 k_\perp}{(2\pi)^2}U(\bm{k}_\perp)e^{-i\bm{k}_\perp \bm{b} - ib_zk\left(1-k_\perp^2 / (2k^2)\right)} \frac{\bm{n}\bm{e}_{\bm{k}\Lambda} }{\left(\frac{Z^2}{a^2}+q^2\right)^2} .
\ee
Such longitudinal shift does not alter the cylindrical symmetry and thus does not affect the OAM of the evolved state.

\begin{figure*}[t]
        \centering
	\begin{subfigure}{0.24\linewidth}
        \center
		\includegraphics[width=\linewidth]{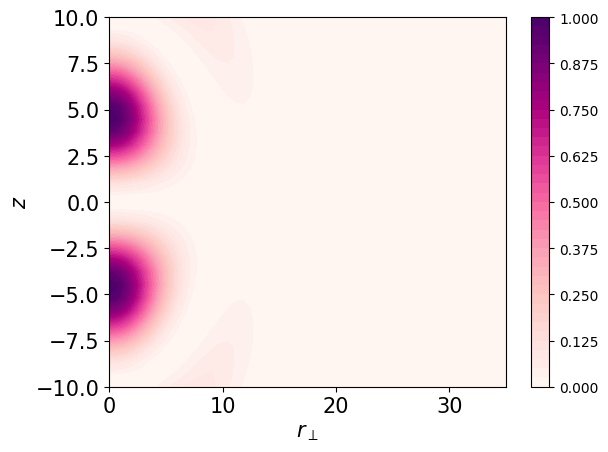}\\(a)
	\end{subfigure}
    \begin{subfigure}{0.24\linewidth}
        \center
		\includegraphics[width=\linewidth]{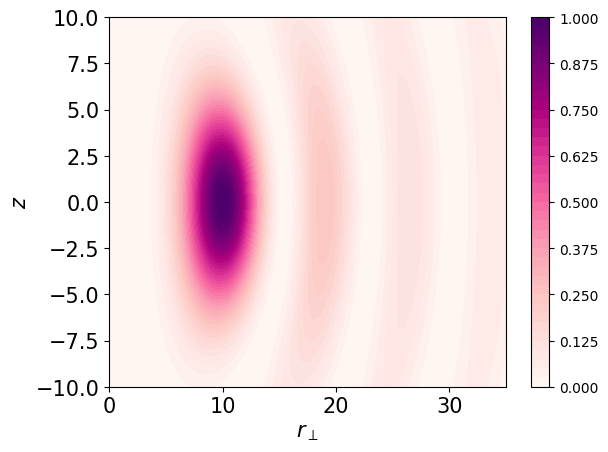}\\(b)
	\end{subfigure}
    \begin{subfigure}{0.24\linewidth}
        \center
		\includegraphics[width=\linewidth]{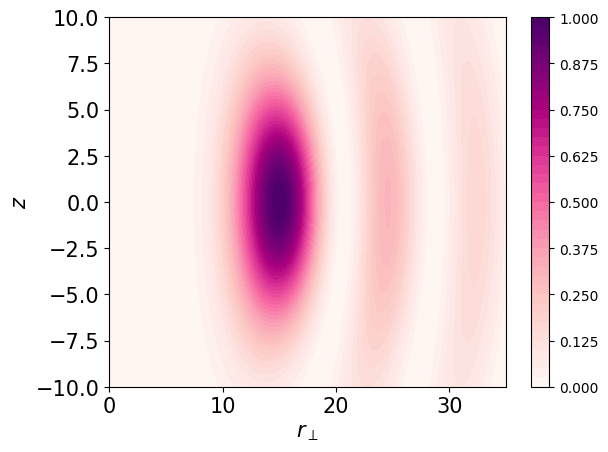}\\(c)
	\end{subfigure}
    \begin{subfigure}{0.24\linewidth}
        \center
		\includegraphics[width=\linewidth]{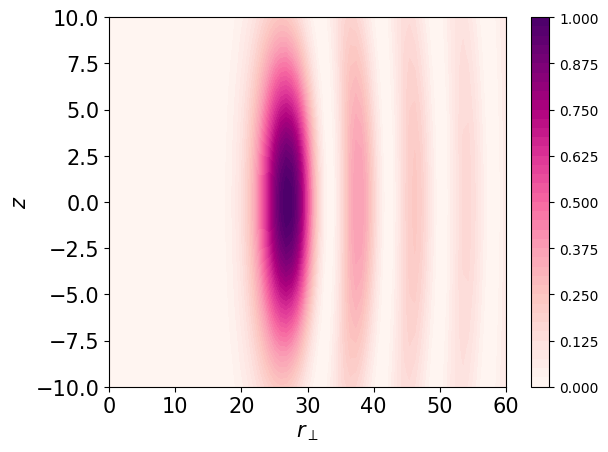}\\(d)
	\end{subfigure}
 
	\caption{Normalized probability density $|\psi_{\text{ev}}(r_\perp, z, \varphi=0)|^2$ of the evolved state for the incident Laguerre-Gaussian beam with $\omega = 1.2|\varepsilon_i|$, $\Lambda$ = 1, $w_0 = 0.5$ mm.
 (a) $l = p = 0$ (Gaussian beam); (b) $\ell = 3, p=1$; (c) $\ell = 5, p=3$; (d) $\ell = 10, p=3$.}
	\label{Prob_dens_Lag}
\end{figure*}

In Fig.~\ref{Prob_dens_Lag}, we show the probability density of the photoelectron for an incident LG beam with different values of the radial index $p$ and OAM $\ell$. The waist $w_0$ is chosen as $0.5$ $\mu$m, approximately equal to the characteristic radius $1/\kappa$ of the Bessel beam for parameters used for these plots. Fig.~\ref{Prob_dens_Lag}~(a) corresponds to a Gaussian mode ($\ell = p = 0$) and does not qualitatively differ from the Bessel beam case with $\ell=0$. Fig. 1 (c) in the main text and Fig. \ref{Prob_dens_Lag} (c) here are also nearly identical. This similarity allows us to conclude in the main text that the width of the photoelectron wave packet is independent of the transverse coherence of the incident light.

\section{Calculating $\expval{\hat L_z}$ with density matrix}
\label{app_mean_L}
Here we check that the average value of the OAM projection upon the $z$ axis calculated with this density matrix equals $\ell$ when the impact parameter equals zero.

Let us start with the definition
\be
    \expval{\hat L_z} = \frac{1}{\Tr(\hat \rho_{\text{ev}})}\Tr\left(\hat L_z \hat \rho_{\text{ev}} \right) = \frac{1}{\Tr(\hat \rho_{\text{ev}})} \int \frac{\dd^3 q}{(2\pi)^3} \bra{\bm q}\hat L_z \hat \rho\ket{\bm q} =\frac{1}{\Tr(\hat \rho_{\text{ev}})}\int \frac{\dd^3 q\dd^3 q'}{(2\pi)^6} \bra{\bm q}\hat L_z \ket{\bm q'}\bra{\bm q'} \hat \rho\ket{\bm q},
\ee
where
\begin{align}
    \bra{\bm q'} \hat \rho\ket{\bm q} 
    =& \psi_{\text{ev}}^*(\bm q) \psi_{\text{ev}}(\bm q'),\\
    \bra{\bm q}\hat L_z \ket{\bm q'} =& \sum_{\tilde \ell} \int \frac{\tilde q_\perp \dd \tilde q_\perp \dd \tilde q_z}{(2\pi)^2} \bra{\bm q}\hat L_z \ket*{\tilde q_\perp \tilde q_z \tilde \ell}\bra*{\tilde q_\perp \tilde q_z \tilde \ell} \ket{\bm q'}.
\end{align}
The projection of a Bessel vortex state on a plane wave being
\begin{equation}
    \bra*{\tilde q_\perp \tilde q_z \tilde \ell} \ket{\bm q} = e^{-i\tilde \ell \varphi_q} \frac{(2\pi)^2}{\sqrt{\tilde q_\perp}} \delta(\tilde q_\perp - q_\perp)\delta(\tilde q_z - q_z),
\end{equation}
we can calculate
\begin{equation}
    \bra{\bm q}\hat L_z \ket{\bm q'} = \sum_{\tilde \ell}\tilde \ell (2\pi)^2 e^{i\tilde \ell (\varphi_q - \varphi_q')} \delta(q_\perp - q_\perp') \delta(q_z -q_z') \frac{1}{q_\perp}.
\end{equation}
Then, with $\dd^3 q=q_\perp \dd  q_\perp \dd  q_z \dd \varphi_q$, we can show
\begin{widetext}
\begin{multline}
    \expval{\hat L_z} = \frac{1}{\Tr(\hat \rho_{\text{ev}})}\int \frac{\dd^3 q\dd^3 q'}{(2\pi)^6} \sum_{\tilde \ell}\tilde \ell (2\pi)^2 e^{i\tilde \ell (\varphi_q - \varphi_q')} \delta(q_\perp - q_\perp') \delta(q_z -q_z') \frac{1}{q_\perp} \psi_{\text{ev}}^*(\bm q) \psi_{\text{ev}}(\bm q')\\
   =\frac{1}{\Tr(\hat \rho_{\text{ev}})}\sum_{\tilde \ell}\tilde \ell \int \frac{q_\perp \dd  q_\perp \dd  q_z \dd \varphi_q \dd \varphi_q'}{(2\pi)^6}  (2\pi)^2 e^{i\tilde \ell (\varphi_q - \varphi_q')} \underbrace{\psi_{\text{ev}}^*(q_\perp, q_z, \varphi_q)}_{\chi^*(q_\perp, q_z)e^{-i \ell \varphi_q}} \underbrace{\psi_{\text{ev}}(q_\perp, q_z,\varphi_q')}_{\chi(q_\perp, q_z)e^{i \ell \varphi_q'}}\\
=\frac{1}{\Tr(\hat \rho_{\text{ev}})}\sum_{\tilde \ell}\tilde \ell \delta_{\ell \tilde \ell} \int \frac{q_\perp \dd  q_\perp \dd  q_z \dd \varphi_q }{(2\pi)^6}  (2\pi)^3 e^{i(\tilde \ell - \ell)\varphi_q}  |\chi(q_\perp, q_z)|^2  \\
 = \frac{1}{\Tr(\hat \rho_{\text{ev}})}\ell \int \frac{q_\perp \dd  q_\perp \dd  q_z \dd \varphi_q }{(2\pi)^3} |\psi_{\text{ev}}(\bm q)|^2 = \ell
\end{multline}
\end{widetext}



\section{Averaging the amplitude over an axially symmetric target}
Let us consider an atomic target described with the two-dimensional distribution function $f(\bm b)$. Assume the target is axially symmetric, which means that $f(\bm b)$ depends only on the absolute value of impact parameter $\bm b$, but not on the azimuthal angle $\varphi_b$.

In the coherent averaging scenario the amplitude of the process  should be averaged over the target:

\begin{equation}
\label{tilde M}
    \widetilde{M}_{fi}^{\text{TW}} = \int b\:\dd b \:\dd \varphi_b f(\bm b) M_{fi}^{\text{TW}}.
\end{equation}

From the amplitude $M_{fi}^{\text{TW}}$ (Eq. (11) in the main text) we recall, first, that dependence on the angle $\varphi_b$ is contained only in the functions
\begin{equation}
    I_{\ell}(\alpha, \beta, \bm{b}, \varphi_p) \equiv \int \frac{\dd \Tilde{\varphi}}{2\pi} e^{i\ell \tilde{\varphi} - i\kappa b \cos (\Tilde{\varphi}+\varphi_p-\varphi_b)}\frac{1}{(\alpha - \beta \cos{\Tilde{\varphi}})^2}
\end{equation}
and, second, that there is a separate phase factor $e^{i\ell \varphi_p}$.
Since the order of integration over $\varphi_b$ and $\tilde \varphi$ does not matter, we can first take the integral over the target and find out which terms are eliminated:

\begin{multline}
\label{d phi_b}
    \int \dd \varphi_b e^{- i\kappa b \cos (\Tilde{\varphi}+\varphi_p-\varphi_b)} = \int \dd \varphi_b \sum_{m=0}^\infty \frac{(-i\kappa b)^m}{m!} \cos^m(\tilde \varphi+ \varphi_p-\varphi_b)  \\ =\int \dd \varphi_b \sum_{m=0}^\infty\left(-\frac{i\kappa b}{2}\right)^m \frac1{m!} \left(e^{i(\tilde \varphi+ \varphi_p-\varphi_b)} + e^{-i(\tilde \varphi+ \varphi_p-\varphi_b)}\right)^m 
     =\int \dd \varphi_b \sum_{m=0}^\infty \frac1{m!} \left(-\frac{i\kappa b}{2}\right)^m \sum_{k=0}^m \binom{m}{k} e^{i(\tilde \varphi+ \varphi_p-\varphi_b)(m-2k)} \\= 2\pi \sum_{m=0}^\infty \frac1{m!} \left(-\frac{i\kappa b}{2}\right)^m \sum_{k=0}^m \binom{m}{k} e^{i(\tilde \varphi+ \varphi_p)(m-2k)}\delta_{m,2k}=2\pi \sum_{m'=0}^\infty \frac1{(2m')!} \binom{2m'}{m'}\left(-\frac{i\kappa b}{2}\right)^{2m'}= {2\pi} J_0(\kappa b) .
\end{multline}
Here $\binom{m}{k}$ are the binomial coefficients. Alternatively, this result can be obtained using the Jacobi–Anger expansion. 

We can conclude that the dependence on the azimuthal angle $\varphi_p$ of the electron momentum in the averaged amplitude \eqref{tilde M} is contained only in the phase factor $e^{i\ell \varphi_p}$, as the final result in Eq. \eqref{d phi_b} contains no phase terms.  That means that thus averaged amplitude is an eigenfunction of the operator $\hat L_z$ just like for a twisted photon interacting with a single atom in a head-on $\bm b=0$ scenario.

\bibliography{ref}